 \documentclass[preprint,12pt]{elsarticle}

 \usepackage{graphicx}

 \usepackage{amssymb}
 \usepackage{paralist}

 \usepackage{amsmath}
 \newtheorem{theorem}{Theorem}
 \newtheorem{corol}{Corollary}
 \newtheorem{lemma}{Lemma}

 \journal{Journal Name}

\begin{document}
 
 \begin{frontmatter}
 
 

 \title{Direct Computing on Control Capability for Linear \\
 Continuous-time Systems Based on Hurwitz Matrix 
 \thanks{Work supported by the National Natural Science Foundation of China (Grant No. 61273005)}}
 
 
 \author{Mingwang Zhao}
 
 \address{Information Science and Engineering School, Wuhan University of Science and Technology, Wuhan, Hubei, 430081, China \\
 Tel.: +86-27-68863897 \\
 Work supported by the National Natural Science Foundation of China (Grant No. 61273005)}
 
 \begin{abstract}
 In this paper, based on the controllable canonical form and the Hurwitz matrix of the Hurwitz stability criterion, an analytical volume computing method for the smooth controllability zonotope for the linear continuous-time(LCT) systems, without of help of the eigenvalue computing of the systems, is presented. And then, the computing method is generlized to the volume computing of the controllability ellipsoid of the LCT systems. Because the controllability zonotope and ellipsoid are directly related to control capability and their volumes are the main index describing the control capability, the new volume computing methods proposed in this paper can help greatly the computing, analysis and optimization of the control capability of LCT systems.
 \end{abstract}
 
 
 \end{frontmatter}
 
 
 \section{Introduction}
 \label{S:1} 
 
 In recent years, based on the concepts and criteria about the state controllability of linear systems, Zhao has done a series of new works on control ability and efficiency of the input variables to the state variables and the state space, and then many novel concepts, analysis and computing methods about that are putforth.
 In fact, the so-called control ability and efficiency, named as the control capability, is a very key concept in the control theory and engineering and can help us to understand, analyze, optimize and design the open-loop contrlled plants, controller and the closed-loop control systems in many aspects. The control capability, so to speak, is a solid foundation of the future delvelopment of the control theory and engineering.
 
 In Zhao's series of works, main contributions are:
 \begin{enumerate}
 \item The controllability zonotope of the linear discrete-time(LDT) systems, that is, the controllability region under unit input variables $ \left( \left \Vert u_t \right \Vert _ \infty \le 1 \right)$, is defined firstly for decribing the control capability with the time-optimality, and then the relations among the control capability of the open-loop controlled plants, control Strategy Space of the controller, and the performance of the closed-loop control systems in paper \cite{zhaomw202003};
 
 \item The concept on the control capability is generlized to the linear continuous-time(LCT) systems and then smooth controllability zonotope for the LCT systems is defined \cite{zhaomw202101}.
 
 \item In papers \cite{zhaomw202001}, \cite{zhaomw202101}, and \cite{zhaomw202201}, many effective volume computing methods for the controllability zonotopes are proven, such as, 
 \begin{compactitem}
 \item two recrusive computing equations for the finite-time controllability zonotope (FTCZ) for LDT systems which all eigenvelues satisfy that $ \lambda_i \in [0,1)$ \cite{zhaomw202001}.
 
 \item the analytical computing equation for the infinite-time controllability zonotope (IFTCZ) for LDT systems which all eigenvelues satisfy that $ \lambda_i \in [0,1)$ \cite{zhaomw202001}.
 
 \item the analytical computing equation for the IFTCZ for LCT systems which all eigenvelues satisfy that $ \lambda_i \in (- \infty,0)$ \cite{zhaomw202101}.
 
 \item the analytical computing equation for the FTCZ for LDT systems and LCT systems which all eigenvelues satisfy that $ \lambda_i \in [0,1)$ \cite{zhaomw202101}.
 \end{compactitem}
 \item The analytical factors descrbing the control capability of the LDT systems are desconstructed from the analytical volume-computing equation of the controllability zonotope and can be very helpful for the analyzing and opyimizing the control capability \cite{zhaomw202004}.
 
 \item The controllability ellipsoids the LDT systems, that is, the controllability region under unit-total-energy input variables $ \left( \int_0^T \left \Vert u_t \right \Vert _2 ^2 \le 1 \right)$, is defined for decribing the control capability with the energy-optimality, and then above works for the controllability capability with the time-optimality are generlized to the control capability with the energy-optimality in paper \cite{zhaomw202002}.
 \end{enumerate}
 In this paer, based on the controllability canonical form (CCF) and the Hurwitz matrix of the Hurwitz stability criterion, a directly analytical volume-computing method for the IFTCZ for the LCT systems and it will greatly improve computing efficiency and will be helpful the computing, anlyzing and opyomizing of the control capability.
 
 \section{Definition on the Control Capability with Time Attribute}
 
In general, the LCT Systems can be modelled as follows:
 \begin{equation}
 \dot x_{t}=Ax_{t}+Bu_{t}, \quad x_{t} \in R^{n},u_{t} \in R^{r}, \label{eq:a220201}
 \end{equation}
 \noindent where $x_{t}$ and $u_{t}$ are the state variable and input
 variable, respectively, and matrices $A \in R^{n \times n}$ and $B \in
 R^{n \times r}$ are the state matrix and input matrix, respectively \cite{Kailath1980} \cite{Chen1998}. 
 To investigate the controllability of the
 LCT systems \eqref{eq:a220201}, the controllability matrix and the controllability Grammian matrix can be defined as follows
 \begin{align}
 P_{n} & = \left[ B, AB, \dots,A^{n-1}B \right ]
 \label{eq:a220202} \\
 G_{T} & = \int _{0} ^{T} e^{At}B \left( e^{At}B \right)^T \textnormal {s} t \label{eq:a220203}
 \end{align} 
 That the rank of the matrix $P_{n}$/$G_{T}$ is $n$, \textit{that is}, the dimension of of the state space the systems \eqref{eq:a220201}, is the well-known criterion on the state controllability.

In papers \cite{zhaomw202001}, \cite {zhaomw202002} and \cite {zhaomw202101}, two smooth geometry generated by the matrix pair $(A,B)$, named as the controllability zonotope and controllability eiilipsoid, can be defined as follows
 \begin{align} 
R_a(T) & = \left \{x : x= \int _{0}^{T} e^{At} Bz_t \textnormal {d} t , \; \; \forall z_t: \Vert z_t \Vert _ \infty \le 1 \right \} \label{eq:a220204} \\
R_b(T) & = \left \{x : x= \int _{0}^{T} e^{At} Bz_t \textnormal {d} t , \; \; \forall z_t: \Vert z_t \Vert _ 2 \le 1 \right \} \label{eq:a220205}
 \end{align}
where the smooth geometry $R_a(T)$ and $R_b(T) $ are a convex $n$-dimensional geometry with the origin symmetry. 

In fact, the controllability regions defined as Eqs. \eqref {eq:a220204} and \eqref{eq:a220205} are the reachability regions and can be regarded as a general term of the reachability regions and the narrow-sense controllability regions. The narrow-sense controllability regions are also named as the recover regions \cite{Hulin2001}
It can be proven that the two kinds of region can be transformed each other by the linear transformation. Therefore, without losing generality, the analysis and volume computing for the broad-sense controllability regions as Eqs. \eqref {eq:a220204} and \eqref{eq:a220205} are studied, and then the obtained results can be easily generlized to the narrow controllability regions.

It is pointed that the above controllability regions, include their volumes, shapes and radii, are directly related to the control capability of the systems $ \Sigma (A,B)$, such as the time-related control capability and the eanergy-related control capability \cite{zhaomw202002}, \cite{zhaomw202003}. Therefore, the volume and shape factors can be used to defined these control capabilities, and then the effective analysis and computing on that are the basis of the analysis and optimizing the these control capabilities for promoting the performance of the open-loop plants or the closed-loop control system. The analytical computing of these volume and the shape factors are studied in papers \cite{zhaomw202004} and some effective computing methods based on the Jordan and diagonal canonical forms, where all eigenvalues, eigenvectors and the generalized eigenvectors are need to be solved with a little computing complexity. 
 
 Next, the volume computings of the IFTCZ based on the CCF and Hurwitz matrix are studied respectively. The new computing methods will no longer depend on the solving of eigenvalues and eigenvectors and are with the lower computational complexity. Later, the ideas deduced the new computing methods will also be generalized to the volume computing of the infite-time controllability ellipsoid (IFTE).

 \subsection {the analytical computing of the volume of the IFTCZ}
 
 In papers \cite{zhaomw202001} and \cite{zhaomw202101}, an analytic volume-computing methods for the samooth zonotope $R_a (T)$ generated by the matrix pair $(A,B)$ with $n$ real eigenvalues of matrix $A$ are proven and then quantifying and maximizing of the controllability region $R_a (T)$ for the LCT systems $ \Sigma(A,B)$ can be carried out based on the results in paper \cite{zhaomw202101}. Because that the computing methods are proven based on the 3 kinds of the Jordan or diagonal canonical forms, there computing equations are with 3 kinds of expressions. 

As we know, any LCT models can be transformed as the Jordan canonical form (the diagonal canonical form can be regarded as a special case of the Jordans). Therefore, for the single-input linear systems $ \Sigma(A,B)$, the Jordan canonical forms $ \Sigma \left( A_J,B_J \right)= \Sigma \left( P_J^{-1}AP,P_J^{-1}B \right) $ with the Jordan transformation matrix $P_J$ are as follows
 \begin{enumerate}
 \item The diagonal canonical form: 
 $ \Sigma \left( \textnormal{diag} \left \{ \lambda_1,  \cdots, \lambda_n \right \} , \left [ \beta_1,\cdots, \beta_n \right ]^T \right)$; 
 where '$ \textnormal{diag}$' means the diagonal matrix.
 
 \item The Jordan canonical form with one Jordan block: 
 $ \Sigma \left( A_{ \lambda} , \left [ \beta_1, \cdots, \beta_n \right ]^T \right)$,
 where $A_{ \lambda}$ is a $n \times n$ Jordan block with the eigenvalue $ \lambda$;
 
 \item The Jordan canonical form with multiple Jordan blocks: 
$$ \Sigma \left( \textnormal{block-diag} \left \{A_1, \cdots,A_q \right \} , \left [B_1^T, \cdots,B_q^T \right ]^T \right)$$
where '$ \textnormal{block-diag}$' means the block diagonal matrix, $A_i$ is a $m_i \times m_i$ Jordan block with the eigenvalue $ \lambda_i$ , $B_i= \left [ \beta_{i,1}, \cdots, \beta_{i,m_i} \right ]^T$, $ \sum_{i=1}^q=n$.
 \end {enumerate}

In fact, the 3-th case of the Jordan canonical forms is a general case, and the 1-st and 2-nd cases are the some special forms of the 3-th case.

 \begin{theorem} \label {th:t220201} If matrix $A \in R^{n \times n}$ is with $n$ different eigenvalues in the interval $(- \infty,0)$ and $b \in R^n$, the volume of the IFTCZ $R_a ( \infty)$ generated by matrix pair $ ( A, B) $ can be computed analytically by the following equation:
 \begin{align} 
V_1 &=2^n \left| \det \left(P_J \right) \left( \prod_{1 \leq i<j \leq n} \frac{ \lambda_{j}- \lambda_{i}}{ \lambda_{i}+ \lambda_{j}} \right) \left( \prod_{i=1}^{n} \frac{ \beta_i}{ \lambda_{i}} \right) \right| \notag \\
 & = 2^n \left| \frac { \det \left(P_J \right) } {L_n \left ( \Lambda_n \right)} \left[ \prod_{1 \leq i<j \leq n} \left ( \lambda_{j}- \lambda_{i} \right ) \right] \left( \prod_{i=1}^{n} \beta_i \right) \right|
 \label{eq:a220206} \\
 V_2 & = 2^n 
 \left \vert \frac{ \det \left (P_J \right) \left( \beta_n/ \lambda \right) ^{n} } { \left( 2 \lambda \right) ^{n(n-1)/2}}
 \right \vert 
 = 2^n \left \vert \frac{ \det \left (P_J \right) \beta_n ^{n} } { L_n \left ( \Lambda_n \right) }
 \right \vert \label{eq:a220207} \\
 V_3 & = 2^n \left \vert \det \left (P_J \right) \prod _{1 \leq i<j \leq q} \left( \frac{ \lambda_j - \lambda_i}{ \lambda_i + \lambda_j } \right)^{m_i \times m_j} \right \vert
 \times 
 \left \vert \prod _{i=1}^{q} \frac{ \left( \beta_{i,m_i}/ \lambda_i \right) ^{m_i} } { \left( 2 \lambda_i \right) ^{m_i(m_i-1)/2}}
 \right \vert \notag \\
 & = 2^n \left \vert \frac { \det \left(P_J \right) } {L_n \left ( \Lambda_n \right)} \prod _{1 \leq i<j \leq q} \left( \lambda_j - \lambda_i \right)^{m_i \times m_j} \right \vert \times 
 \left \vert \prod _{i=1}^{q} \beta_{i,m_i} ^{m_i} 
 \right \vert \label{eq:a220208}
 \end{align}
 where $ L_n \left( \Lambda_n \right)$ is a $[n(n+1)/2]$-order homogeneous polynomials on all $n$ eigenvalues $ \Lambda_n= \left [
 \lambda_1, \lambda_2, \cdots, \lambda_n \right]$ defined as follows
 \begin{equation}
 L_n \left ( \Lambda_n \right)= \left[ \prod_{1 \leq i<j \leq n} \left( \lambda_{i}+ \lambda_{j} \right) \right] \left( \prod_{i=1}^{n} \lambda_{i} \right) \label{eq:a220209}
 \end{equation} 
where $ \Lambda_n$ includes all eigenvalues and their multiplicities.
 \end{theorem}

 Based on the above theorems, the volume of the controllability region $R_a (N)$ when $N \rightarrow \infty$ can be computed analytically.
 Furthermore, some analytical factors describing the shape of the $R_a (N)$ can be got by deconstructing the analytical volume computing equation \eqref{eq:a220206}. These analytical expressions for these volume and shape factors can be describe quantitatively the control capability of the dynamical systems, and then optimizing these volume and shape factors is indeed maximizing the control capability. 
 
 \section {the direct computing of the IFTCZ based on Hurwitz matrix}
 \subsection {The analytical volume expression of of the IFTCZ of the systems $ \Sigma \left(A_c,B_c \right)$ }
Considering the following controllable canonical form (CCF) of the single-input LCT systems 
 \begin{align}
 \dot x_c&=A_cx_c+B_cu \notag \\
& =
 \left[ \begin{array}{cccc}
 0 & 1 & \cdots & 0 \\
 \vdots & \vdots & \ddots & \vdots \\
 0 & 0 & \cdots & 1 \\
 -a_{n} & -a_{n-1} & \cdots & -a_{1}
 \end{array} \right]x_c+ \left[ \begin{array}{c}
 0 \\
 \vdots \\
 0 \\
 1
 \end{array} \right]u \label{eq:a220210}
 \end{align}

In fact, by the linear system theory, any state controllable LCT systems 

$$ \Sigma(A,B): \, x=Ax+Bu, $$
 \noindent can be transformaed as the above CCF $ \Sigma \left(A_c,B_c \right)$ by the state transformation $x=W_cx_c$, where the transformation matrix $W_c$ \cite{Kailath1980} \cite{Chen1998}

 \begin{align}
 W_c&= \left[ \begin{array}{c}w_1 \\ w_1A \\ \vdots \\w_1A^{n-1} \end{array}
 \right]^{-1} \label {eq:a220211}\\
 w_1&=[0,0, \dots,1]P_n^{-1} \notag \\
 &=[0,0, \dots,1] \left[ B, AB, \dots,A^{n-1}B \right ]^{-1} \label {eq:a220212}
 \end{align}

By the \textbf{Theorem \ref{th:t220201}} for volume computing of the IFTCZ, we have the following corollary for volume computing of CCF

 \begin{corol} \label {cor:c220201} If the matrix $A_c$ of the CCF \eqref{eq:a220210} is with $n$ eigenvalues in the interval $(- \infty,0)$ and $B_c \in R^n$, the volume of the IFTCZ $R_a ( \infty)$ generated by matrix pair $ \Sigma \left(A_c,B_c \right)$ can be computed analytically by the following equation:
 \begin{equation} \label{eq:a220213}
 \textnormal {vol} (R_ \infty) = \left| \frac {2^n} { L_n \left ( \Lambda_n \right)} \right| 
 \end{equation}
And then, for any controllable LCT systems $ \Sigma(A,B)$, the volume of the correponding IFTCZ can be computed analytically as follows
 \begin{equation} \label{eq:a220240}
	 \textnormal {vol} (R_ \infty) = \left| \frac {2^n \det \left (W_c \right)} { L_n \left ( \Lambda_n \right)} \right| 
 \end{equation} \end{corol}

 \textbf{Proof} The corollary will be prven in 3 cases of the Jordan canonical forms as follows.
 
 1) Firstly, the case that all $n$ eigenvalues of the matrix $A$ are differential is considered.
 
 For the matrix $A_c$ of the CCF \eqref{eq:a220210}, the eigenvector $v_i$ corresponding to the eigenvalue $ \lambda_i$ can be taken as 
 \begin{align} 
 v_i= \left[ 1 \; \lambda_i \; \lambda_i^2 \; \cdots \; \lambda_i^{n-1} \right]^T \label{eq:a220215}
 \end{align}
And then, the diagonalization transformation of the systems $ \Sigma \left(A_c,B_c \right)$ can be choosen as 
 \begin{align} 
 P_J= \left[ v_1 \; v_2 \; \cdots \; v_n \right] \label{eq:a220216}
 \end{align}
 and the corresponding diagonal canonical form is
 \begin{align}
 \dot { \tilde x} &=P_J^{-1} A_c P_J \tilde x+ P_J^{-1} B_c u \notag \\
 & =
 \left[ \begin{array}{cccc}
 \lambda_1 & 0 & \cdots & 0 \\
 0 & \lambda_2 & \cdots & 0 \\
 \vdots & \vdots & \ddots & \vdots \\
 0 & 0 & \cdots & \lambda_n
 \end{array} \right]x+ \left[ \begin{array}{c}
 \beta_1 \\
 \beta_2 \\
 \vdots \\
 \beta_n
 \end{array} \right]u \label{eq:a220217}
 \end{align}
 \noindent where

 \begin{align}
 \det \left(P_J \right) & = \prod_{1 \le i<j \le n} \left( \lambda_{j}- \lambda_{i} \right) \label{eq:a220218} \\
P_J^{-1} &
= \frac{ \textrm{adj} \left(P_J \right)}{ \left|P_J \right|} 
= \left[ \begin{array}{cccc}
 * & \cdots & * & \beta_{1} \\
 * & \cdots & * & \beta_{2} \\
 \vdots & \ddots & \vdots & \vdots \\
 * & \cdots & * & \beta_{n}
 \end{array} \right] \label{eq:a220219} \\
 \beta_{i} &=(-1)^{n-1} \prod_{ j\in \{1,n\} \setminus i} \frac{1}{ \lambda_{j}- \lambda_{i}} \; i=1,2, \cdots,n \label{eq:a220220}
 \end{align}
where '$\{1,n\} \setminus j$' means the set $\{1,2,\cdots,n\}$ but doesn't include number $j$.

By \textbf {Theorem \ref {th:t220201}}, the volume of the IFTCZ zonotope $R_a ( \infty)$ generated by matrix pair $ \Sigma \left(A_c,B_c \right)$ is
 \begin{align}
 \textrm{Vol} \left(E_{n} \right)
 &= \left| \det \left(P_J \right) \left( \prod_{1 \le i<j \le n} \frac{ \lambda_{j}- \lambda_{i}}{ \lambda_{j}+ \lambda_{i}} \right) \left( \prod_{i=1}^{n} \frac{ \beta_{i}}{ \lambda_{i}} \right) \right| \notag \\
 &= \left| \prod_{1 \le i<j \le n} \left( \lambda_{j}- \lambda_{i} \right) \left( \prod_{1 \le i<j \le n} \frac{ \lambda_{j}- \lambda_{i}}{ \lambda_{j}+ \lambda_{i}} \right) \left( \prod_{i=1}^{n} \frac{(-1)^{n-1}}{ \lambda_{i}} \prod_{ j \in \{1,n\} \setminus i} \frac{1}{ \lambda_{j}- \lambda_{i}} \right) \right| \notag \\
&= \left| \left( \prod_{1 \le i<j \le n} \frac{ \left( \lambda_{j}- \lambda_{i} \right)^{2}}{ \lambda_{j}+ \lambda_{i}} \right) \left( \prod_{i=1}^{n} \frac{1}{ \lambda_{i}} \right) \left( \prod_{1 \le j \le n} \frac{1}{ \left( \lambda_{j}- \lambda_{i} \right)^{2}} \right) \right| \notag \\
&= \left| \left( \prod_{1 \le i<j \le n} \frac{1}{ \lambda_{j}+ \lambda_{i}} \right) \left( \prod_{i=1}^{n} \frac{1}{ \lambda_{i}} \right) \right| \label{eq:a220221}
 \end{align}

2) Secondly, the case that all $n$ eigenvalues of the matrix $A$ are same as $ \lambda$, that is, the Jordan matrix correponding to the matrix $A$ is with on Jordan block.

For the matrix $A_c$ of the CCF \eqref{eq:a220217}, the eigenvector and generalized eigenvector corresponding to the eigenvalue $ \lambda$ can be taken as 
 \begin{align} 
	v_1&= \left[ 1 \; \lambda \; \lambda^2 \; \cdots \; \lambda^{n-1} \right]^T \notag \\
		v_2&= \left[ 0 \; 1 \; 2 \lambda \; \cdots \; (n-1) \lambda^{n-2} \right]^T \notag \\
		v_3&= \left[ 0 \; 0 \; 1 \; \cdots \; \frac {(n-1)!}{2!(n-3)!} \lambda^{n-3} \right]^T \notag \\
		& \cdots \cdots \notag \\
		v_n&= \left[ 0 \; 0 \; 0 \; \cdots \; 1 \right]^T \label{eq:a220222}
 \end{align}
And then, the diagonalization transformation of the systems $ \Sigma \left(A_c,B_c \right)$ can be choosen as 
 \begin{align} 
	P_J&= \left[ v_1 \; v_2 \; \cdots \; v_n \right] \notag \\
	&= \left[ \begin{array}{cccccc}
		1 & 0 & 0 & \cdots & 0 \\
		 \lambda & 1 & 0 & \cdots & 0 \\
		 \lambda^2 & 2 \lambda & 1 & \cdots & 0 \\
		 \vdots & \vdots & \vdots & \ddots & \vdots \\
		 \lambda^{n-1} & (n-1) \lambda^{n-2} & \frac{(n-1)!}{2!(n-3)!}	 \lambda^{n-3} & \cdots & 1 
		 \end{array}
	 \right] \label{eq:a220223}
 \end{align}
and the corresponding Jordan canonical form is
 \begin{align}
	 \dot { \tilde x} & =
	 \left[ \begin{array}{cccc}
		 \lambda & 1 & \cdots & 0 \\
		0 & \lambda & \cdots & 0 \\
		 \vdots & \vdots & \ddots & \vdots \\
		0 & 0 & \cdots & \lambda
	 \end{array} \right]x+ \left[ \begin{array}{c}
		 \beta_1 \\
		 \beta_2 \\
		 \vdots \\
		 \beta_n
	 \end{array} \right]u \label{eq:a220224}
 \end{align}
 \noindent where

 \begin{align}
	 \det \left(P_J \right) & = 1 \notag \\
	P_J^{-1} &
	= \frac{ \textrm{adj} \left(P_J \right)}{ \left|P_J \right|} 
	= \left[ \begin{array}{cccc}
		* & \cdots & * & \beta_{1} \\
		* & \cdots & * & \beta_{2} \\
		 \vdots & \ddots & \vdots & \vdots \\
		* & \cdots & * & \beta_{n}
	 \end{array} \right] \notag  \\
	 \beta_{i} &= \left \{ \begin{array}{ll}
		0 & i < n \\
		1 & i=n
		 \end{array}
	 \right. \notag 
 \end{align}

By \textbf {Theorem \ref {th:t220201}}, the volume of the IFTCZ $R_a ( \infty)$ generated by matrix pair $ \Sigma \left(A_c,B_c \right)$ is
 \begin{align}
	 \textrm{Vol} \left(E_{n} \right)
	&= 2^n \left \vert \frac{ \det \left (P_J \right) \beta_n ^{n} } { L_n \left ( \Lambda_n \right) }
	 \right \vert = \left \vert \frac{ 2^n} { L_n \left ( \Lambda_n \right) }
 \right| \label {eq:a220225}
 \end{align}

3) Combining above proving processes for the two kind of Jordan canonical forms, it can be proven that, Eq. \eqref{eq:a220213} for the CCF \eqref{eq:a220224} which the corresponding Jordan matrix is with the multiply Jordan blocks is also true.

To sum up, when all eigenvalues the CCF \eqref{eq:a220224} are negative numbers, whether they are single or repeated eigenvalues, Eq. \eqref{eq:a220213} for the volume computing of the IFTCZ generated by the matrix pair $ \Sigma \left( A_c,B_c \right)$ holds.
 \subsection {The determinant of the Hurwitz matrix}

Consider the following $n$-order real polynomial $ f^{(n)}(s) $ with the zeros $ \left \{ \lambda_1, \lambda_2, \cdots, \lambda_n \right \}$
 \begin{align}
 f^{(n)}(s) 
 & =s^n+a^{(n)}_1s+ \cdots+a^{(n)}_{n-1}s +a^{(n)}_{n} \notag \\
 &= \prod _{i=1}^{n} \left(s- \lambda _i \right) \label {eq:a220226}
 \end{align}
 where 
 \begin{align}
 a^{(n)}_k = (-1)^k \sum_{1 \le i_{1}< \cdots<i_{k} \le n} \prod_{j=1}^{k} \lambda_{i_{j}} \label {eq:a220227}
 \end{align} 
 \qed

For the $n$-order real polynomial $f^{(n)}(s)$, the correponding Hurwitz matrx can be defined as follows

 \begin{align}
H^{(n)} = \left[ \begin{array}{cccccccc}
 a_{1}^{(n)} & a_{3}^{(n)} & a_{5}^{(n)} & a_{7}^{(n)} & \cdots & 0 & 0 & 0 \\
 1 & a_{2}^{(n)} & a_{4}^{(n)} & a_{6}^{(n)} & \cdots & 0 & 0 & 0 \\
 0 & a_{1}^{(n)} & a_{3}^{(n)} & a_{5}^{(n)} & \cdots & 0 & 0 & 0 \\
 0 & 1 & a_{2}^{(n)} & a_{4}^{(n)} & \cdots & 0 & 0 & 0 \\
 \vdots & \vdots & \vdots & \vdots & \ddots & \vdots & \vdots & \vdots \\
 0 & 0 & 0 & 0 & \cdots & a_{n-2}^{(n)} & a_{n}^{(n)} & 0 \\
 0 & 0 & 0 & 0 & \cdots & a_{n-3}^{(n)} & a_{n-1}^{(n)} & 0 \\
 0 & 0 & 0 & 0 & \cdots & a_{n-4}^{(n)} & a_{n-2}^{(n)} & a_{n}^{(n)}
 \end{array} \right] \label {eq:a220228}
 \end{align}
 \noindent By Eq. \eqref {eq:a220227} for computing coefficients $a^{(n)}_k$, we see, the deteminant value of the Hurwitz matrix $H^{(n)}$ is a $[n(n+1)/2]$-order homogeneous polynomial about the variabl vector $ \Lambda_n$.


By the Hurwitz criterion, we know, the conditions that the all zeros of the polynomial $f^{(n)}(s)$ are with the negative real parts, that is, all poles of the LCT systems are with the negative real parts, and then systems are stable is the following order principal minor determinants of the Hurwitz matrix $H^{(n)}$ are greater than 0, that is,

 \begin{align*}
 H^{(n)}_i>0, \, i=1,2, \cdots,n
 \end{align*}
 \noindent where the order principal minor determinant $H^{(n)}_i$ is the $i \times i$ matrix composed by the first $i$ rows and first $i$ columns of the matrix 
$H^{(n)}$.


For the deteminant of the Hurwitz matrix $H^{(n)}$, we have thw following lemma.

 \begin{lemma} \label{lem:L220201}
	For the $n$-order real polynomial $f^{(n)}(s)$, the determinat of the correponding Hurwitz matrix $H^{(n)}$ can be computed as 
 \begin{align}
 \det \left(H^{(n)} \right) &= L_n \left( \Lambda_n \right) = \left( \prod_{1 \le i<j \le n} \lambda_{j}+ \lambda_{i} \right) \left( \prod_{i=1}^{n} \lambda_{i} \right) \label {eq:a220229}
 \end{align}
 \end{lemma}

 \textbf{Proof.} 
By Eq. \eqref{eq:a220228}, we have 
 \begin{align}
 \det \left(H^{(n)} \right) & = \det \left(H_{n-1}^{(n)} \right) a_{n}^{(n)} \notag = \det \left(H_{n-1}^{(n)} \right) \left( \prod_{i=1}^{n} \lambda_{i} \right)
\notag
 \end{align}

Next, it is proved fistly that for any different $i $ and $j $, $ \lambda_{j}+ \lambda_{i}$ is a factor of $ \det H_{n-1}^{(n)}$. For that, we have the following computing result
 \begin{align}
 \left.a_{k}^{(n)} \right|_{ \lambda_{n-1}=- \lambda_{n}}&=(-1)^k \left. \sum_{1 \le i_{1}< \cdots<i_{k} \le n} \prod_{j=1}^{k} \lambda_{i_{j}} \right|_{ \lambda_{n-1}=- \lambda_{n}} \notag \\
 & = \left \{ \begin{array}{ll}
 a_{1}^{(n-2)} & k=1 \\
 a_{2}^{(n-2)}- \lambda_{n}^{2} & k=2 \\
 a_{k}^{(n-2)}- \lambda_{n}^{2}a_{k-2}^{(n-2)} & k \in \{3,n-2 \} \\
 - \lambda_{n}^{2}a_{k-2}^{(n-2)} & k=n,n-1
 \end{array} \right. \label {eq:a220230}
 \end{align}
where $ \left \{a_{k}^{(n-2)}, k=1,2, \dots,n-2 \right \}$ are the coefficients of the $(n-2)$-order real polynomial $ f^{(n-2)}(s) $ with the zeros $ \left \{ \lambda_1, \lambda_2, \cdots, \lambda_{n-2} \right \}$. Based on that, we have
 \begin{align}
 \left. H_{n-1}^{(n)} \right|_{ \lambda_{n-1}=- \lambda_{n}} & = \left[ \begin{array}{cccccc}
 a_{1}^{(n-2)} & a_{3}^{(n-2)} - \lambda_n^2a_{1}^{(n-2)} & a_{5}^{(n-2)} - \lambda_n^2a_{3}^{(n-2)} & \cdots \\
 1 & a_{2}^{(n-2)} - \lambda_n^2 & a_{4}^{(n-2)}- \lambda_n^2a_{2}^{(n-2)} & \cdots \\
 0 & a_{1}^{(n-2)} & a_{3}^{(n-2)} - \lambda_n^2a_{1}^{(n-2)} & \cdots \\
 0 & 1 & a_{2}^{(n-2)}- \lambda_n^2 & \cdots \\
 \vdots & \vdots & \vdots & \ddots \\
 0 & 0 & 0 & \cdots \\
 0 & 0 & 0 & \cdots 
 \end{array} \right. \notag \\
& \; \; \; \; \; \left. \begin{array}{cccccc}
 0 & 0 \\
 0 & 0 \\
 0 & 0 \\
 0 & 0 \\
 \vdots & \vdots \\
 a_{n-2}^{(n-2)} - \lambda_n^2a_{n-4}^{(n-2)} & - \lambda_n^2a_{n-2}^{(n-2)} \\
a_{n-3}^{(n-2)} - \lambda_n^2a_{n-5}^{(n-2)} & - \lambda_n^2a_{n-3}^{(n-2)} 
 \end{array} \right] \notag \\
& = \left[ \begin{array}{ccccccc}
 a_{1}^{(n-2)} & a_{3}^{(n-2)} & a_{5}^{(n-2)} & \cdots & 0 & 0 \\
 1 & a_{2}^{(n-2)} & a_{4}^{(n-2)}& \cdots & 0 & 0 \\
 0 & a_{1}^{(n-2)} & a_{3}^{(n-2)} & \cdots & 0 & 0 \\
 0 & 1 & a_{2}^{(n-2)} & \cdots & 0 & 0 \\
 \vdots & \vdots & \vdots & \ddots & \vdots & \vdots \\
 0 & 0 & 0 & \cdots& a_{n-2}^{(n-2)} & 0 \\
 0 & 0 & 0 & \cdots & a_{n-3}^{(n-2)} & 0 \\
 \end{array} \right] \label {eq:a220231}
 \end{align}
Therefore, we have
 \begin{align}
 \det \left( \left. H_{n-1}^{(n)} \right|_{ \lambda_{n-1}=- \lambda_{n}} \right)=0 \label {eq:a220232}
 \end{align}
that is, $ \lambda_{n-1}+ \lambda_n$ is a factor of $ \det H_{n-1}^{(n)} $. Without losing generality, it can be summarized as that for any $i$ and $j$, $ \lambda_i+ \lambda_j$ is a factor of $ \det H_{n-1}^{(n)} $. Hence, we have
 \begin{align}
 \det H_{n-1}^{(n)}& = h \left( \Lambda_n \right) \left( \prod_{1 \le i<j \le n} \lambda_{j}+ \lambda_{i} \right) \notag \\
 \det H^{(n)}& = h \left( \Lambda_n \right) \left( \prod_{1 \le i<j \le n} \lambda_{j}+ \lambda_{i} \right) \left( \prod_{i=1}^{n} \lambda_{i} \right)=h \left( \Lambda_n \right)L_n \left( \Lambda_n \right) \label {eq:a220233}
 \end{align}
where $h \left( \Lambda_n \right)$ is an undetermined function on vector $ \Lambda_n$. By Eqs. \eqref{eq:a220228} and \eqref{eq:a220209}, we know,
$ \det H^{(n)} $ and $ L_n \left( \Lambda_n \right)$ are the $[n(n+1)/2]$-order homogeneous polynomials on the vector $ \Lambda_n$. After anylysing these two homogeneous polynomials, we can conclude that $ \lambda_{1}^{n} \lambda_{2}^{n-1} \dots \lambda_{n}$ is a common term of these two polynomials and correspondingly two coefficients of these terms 
 are samed as one. Therefore, we have
 $$h \left( \Lambda_n \right)=1,$$
 and then the lemma is proven true.
 \qed
 \\

Combining the above \textbf{Corollary \ref {cor:c220201}} and \textbf{Lemma \ref{lem:L220201}}, we have the following result about the volume of the IFTCZ based on the Hurwita matrix.


 \begin{theorem} \label{th:t220202}
	For the single-input LCT systems with the state controllability, if all eigenvalues of the systems are the negative real numbers, the volume of the IFTCZ can be computed analytically as follows
	
 \begin{equation} \label{eq:a220234}
 \textnormal {vol} (R_ \infty) = 2^n \left| \frac { \det W_c} { \det H^{(n)} } \right| 
 \end{equation}
 \noindent where the matrix $ W_c$ is the transformation matrix which can transformate the system models $ \Sigma(A,B)$ as the CCF $ \Sigma \left(A_c,B_c \right)$.

 \end{theorem}

To sum up, by now we have 3 kinds of the analytical volume computing methods of the IFTCZ for the LCT systems $ \sigma(A,B)$, presented in \textbf{Theorem \ref{th:t220201}}, \textbf{Corollary \ref{cor:c220201}} and \textbf{Theorem \ref{th:t220202}}. 
The difference among the 3 kind computing methods are summaried as follows:
 \begin{compactitem}
	 \item The all eigenvalues, eignevectors and generalized eigenvectors of the matrix $A$ are needed to be solved out in the methods presented in \textbf{Theorem \ref{th:t220201}} and is with slightly larger computational complexity. In addition, some combination calculation for the matrix $A$ with the corresponding multiply Jordan blocks is with a little trouble;

	 \item The all eigenvalues of the matrix $A$ are needed to be solved out in the method in \textbf{Corollary \ref{cor:c220201}};

	 \item The method in \textbf{Theorem \ref{th:t220202}} is only based on the determinant calculation of Hurwitz matrix and is with less computational complexity.
 \end{compactitem}

 \section {the analytical computing of the volume of the IFTCE}
 
 The above volume computing methods about the IFTCZ based on the CCF and the Hurwita matrix will be generlized to the volume computing of the IFTCE.
 
 For the volume computing problem of the IFTCE of the LCT systems, we have the following theorem about the analytical computing \cite{zhaomw202101}.
 
 \begin{theorem} \label {th:t220203} If the $n$ complex eigenvalues of the matrix $A \in R^{n \times n}$ is with negative real parts and $b \in R^n$, the volume of the IFTCE $R_b ( \infty)$ generated by matrix pair $ ( A, B) $ can be computed analytically by the following equation:
 	 \begin{align} 
 		V_1 &=\Pi_n \left| \det \left(P_J \right) \left( \prod_{1 \leq i<j \leq n} \frac{ \lambda_{j}- \lambda_{i}}{ \lambda_{i}+ \lambda_{j}} \right) \left( \prod_{i=1}^{n} \frac{ \beta_i}{ \left( 2 \lambda_{i} \right) ^{1/2}} \right) \right| \label{eq:a220235}\\
 		V_2 & = \Pi_n 
 		 \left \vert \frac{ \det \left (P_J \right) \beta_n^{n} } { \left( 2 \lambda \right) ^{n^2/2}}
 		 \right \vert 
 		 \label{eq:a220236} \\
 		V_3 & = \Pi_n \left \vert \det \left (P_J \right) \prod _{1 \leq i<j \leq q} \left( \frac{ \lambda_j - \lambda_i}{ \lambda_i + \lambda_j } \right)^{m_i \times m_j} \right \vert
 		 \times 
 		 \left \vert \prod _{i=1}^{q} \frac{ \beta_{i,m_i} ^{m_i} } { \left( 2 \lambda_i \right) ^{m_i^2/2}}
 		 \right \vert \label{eq:a220237}
 	 \end{align}
 	 \noindent where the hypersphere volume-coefficient $\Pi_n$ and the Gamma function $ \varGamma (s)$ can be defined respectively as 
 	 \begin{align} 
 		\Pi_n & = \frac{ \pi^{n/2} } { \varGamma \left( \frac {n}{2}+1 \right)}
 		 \label{eq:a220238} \\
 		 \varGamma(s)
 		& = \left \{ \begin{array}{ll} 
 			(s-1) \varGamma(s-1) & s>1 \\
 			 \sqrt \pi & s=1/2 
 		 \end{array} \right. \label{eq:a220239}
 	 \end{align}
 \end{theorem}

 Based on the above theorems, the volume of the IFTCE $R_b (\infty)$  can be computed analytically.
 Furthermore, some analytical factors describing the shape of the $R_b (\infty)$ can be got by deconstructing the analytical volume computing equations \eqref{eq:a220235}, \eqref{eq:a220236} and \eqref{eq:a220237}. These analytical expressions for these volume and shape factors can be describe quantitatively the control capability of the dynamical systems, and then optimizing these volume and shape factors is indeed maximizing the control capability. 
 
Similar to \textbf{Corollery \ref {cor:c220201} } and \textbf{Theorem \ref{th:t220202}}, we have the following analytical computing corollary and theorem based on the CCF and the Hurwita matrix.
 
 \begin{corol} \label {cor:c220202} 
For any controllable single-input LCT systems $ \Sigma(A,B)$, if all eigenvalues of the matrix $A$ are the negative real numbers, the volume of the correponding IFTCE can be computed analytically as follows
	 \begin{equation} \label{eq:a220240}
		 \textnormal {vol} (R_ \infty) = \Pi_n \left( \frac{a_n}{2^n} \right ) ^{1/2} \left| \frac { \det \left (W_c \right)} { L_n \left ( \Lambda_n \right)} \right| 
	 \end{equation}
	where the matrix $ W_c$ is the transformation matrix which can transformate the system models $ \Sigma(A,B)$ as the CCF $ \Sigma \left(A_c,B_c \right)$; $a_n$ is a constant term of the characteristic polynomial of the matrix $A$ and can be represented as follows 
	$$a_n=(-1)^n \prod_{i=1}^{n} \lambda_{i}$$
 \end{corol}

 \begin{theorem} \label{th:t220204}
	For the single-input LCT systems with the state controllability, if all eigenvalues of the systems are the negative real numbers, the volume of the IFTCZ can be computed analytically as follows
	 \begin{equation} \label{eq:a220241}
		 \textnormal {vol} (R_ \infty) =\Pi_n \left( \frac{a_n}{2^n} \right ) ^{1/2} \left| \frac { \det W_c} { \det H^{(n)} } \right| 
	 \end{equation}
 \end{theorem}

To sum up, by now we have also 3 kinds of the analytical volume computing methods of the IFTCE for the LCT systems $ \Sigma(A,B)$, presented in \textbf{Theorem \ref{th:t220203}}, \textbf{Corollary \ref{cor:c220202}} and \textbf{Theorem \ref{th:t220204}}. 
The difference among the 3 kind computing methods are summaried as follows:
\begin{compactitem}
	\item The all eigenvalues, eignevectors and generalized eigenvectors of the matrix $A$ are needed to be solved out in the methods presented in \textbf{Theorem \ref{th:t220203}} and is with slightly larger computational complexity. In addition, some combination calculation for the matrix $A$ with the corresponding multiply Jordan blocks is with a little trouble;
	
	\item The all eigenvalues of the matrix $A$ are needed to be solved out in the method in \textbf{Corollary \ref{cor:c220202}};
	
	\item The method in \textbf{Theorem \ref{th:t220202}} is only based on the determinant calculation of Hurwitz matrix and is with less computational complexity.
\end{compactitem}
 
 \section {Numerical Experiments (Not available here)}

 \section{Conclusions (Not available here)}

 \bibliographystyle{model1b-num-names}
 \bibliography{zzz}
 \end{document}